\documentclass[12pt,english]{paper}
\usepackage[T1]{fontenc}
\usepackage[latin1]{inputenc}
\usepackage{geometry}
\usepackage{graphicx}
\usepackage{setspace}
\usepackage[numbers]{natbib}

\makeatletter

 \newcommand{\lyxaddress}[1]{
   \par {\raggedright #1 
   \vspace{1.4em}
   \noindent\par}
 }

\usepackage{babel}
\makeatother
\begin{document}

\title{Enhanced spin accumulation in a superconductor}

\author{M. Urech, J. Johansson, N. Poli, V. Korenivski and D. B. Haviland}

\maketitle
\lyxaddress{Nanostructure Physics, Royal Institute of Technology, 10691 Stockholm,
Sweden}
\begin{abstract}
A lateral array of ferromagnetic tunnel junctions is used to inject
and detect non-equilibrium quasi-particle spin distribution in a superconducting
strip made of Al. The strip width and thickness is kept below the
quasi particle spin diffusion length in Al. Non-local measurements
in multiple parallel and antiparallel magnetic states of the detectors
are used to in-situ determine the quasi-particle spin diffusion length.
A very large increase in the spin accumulation in the superconducting
state compared to that in the normal state is observed and is attributed
to a diminishing of the  quasi-particle population by opening of the gap below 
the transition temperature.
\end{abstract}
\newpage
Studies of quasiparticle (QP) injection into superconductors started
with the pioneering experiments by Clarke \cite{Clarke:72}, who
used a thin stack of two normal metal electrodes (N) separated from
a superconducting film (S) by tunnel junctions to create and detect
a charge imbalance of quasiparticles in S. The resulting non-equilibrium
chemical potential in S produced a voltage at the detector junction.
This non-equilibrium voltage was measured with a ``non-local''
technique, where the detector is placed outside the current path (region
of zero charge flow) but sufficiently close to the injection point
for improved sensitivity. Johnson \cite{Johnson:94} used a similar
non-local technique with magnetic injector and detector electrodes
and metallic contacts to measure the spin relaxation parameters in
Nb. The spin relaxation length in superconducting Nb was also studied
using the conventional ``local'' technique \cite{Gu:02}, where
the thin superconductor is enclosed by two ferromagnetic (F) layers
in a spin-valve configuration, and a much smaller temperature variation
of the spin relaxation length compared to that in ref. \cite{Johnson:94}
was found. An indirect measurement of a finite voltage drop over a certain 
length near the interface under spin injection \cite{Shin:03}, indicated a 
decreasing spin diffusion length with temperature. All of these spin transport
measurements on S's, performed with metallic contacts, are sensitive to the 
evanescent wave penetration in S, where the Andreev reflection dominates the
transport \cite{Andreev:64}. In this work we report on spin injection
in F-S structures with the injection performed through tunnel junctions
and a non-local spin-sensitive detection. We use a multi terminal
device, allowing for in-situ determination of the spin diffusion length, which
significantly reduces the sample to sample irreproducibilities. We find a much 
larger spin accumulation, orders of magnitude stronger in S compared to the normal 
metal state, and a small change in the spin relaxation length across the N-S 
transition.

Due to a shift in the density of states at the Fermi level in F, a
net spin asymmetry of conduction is present and can be expressed as
a spin current density, $j_{s}\equiv j_{\uparrow}-j_{\downarrow}\equiv\gamma(j_{\uparrow}+j_{\downarrow})\equiv\gamma j_{q}$,
where $j_{\uparrow(\downarrow)}$ is the current density for carriers
with spin $\uparrow(\downarrow)$, $\gamma$ the spin polarization,
and $j_{q}$ is the charge current. $\gamma$ for Co-based contacts
of $10-40\ \%$ has been reported \cite{Tedrow:73,Johnson:85,Jedema:02,Urech:04,Valenzuela:05}.
Injection of such spin polarised current into a non magnetic metal
(N) creates a spin splitting of the chemical potential at the injection
point, $\delta\mu=\mu_{\uparrow}-\mu_{\downarrow}$, directly determined
by the injected current, $\nabla\delta\mu=e\rho_{N}\gamma\mathbf{j}_{q}$,
with $\rho_{N}$ being the resistivity. The non-equilibrium spin population
diffuses away from the injection point, obeying the diffusion equation,
\begin{equation}
\nabla^{2}\delta\mu=\frac{1}{\lambda_{N}^{2}}\delta\mu,\end{equation}
 where $\lambda_{N}=\sqrt{D_{N}\tau_{N}}$ is the spin diffusion length,
and $D_{N}$ and $\tau_{N}$ are the diffusion constant and spin flip
time in N, respectively. For a 1D wire, having thickness and width
much smaller than the characteristic spin relaxation length, $h< w \ll \lambda_{N}$,
the spin splitting at the interface is $\delta\mu=2e\gamma R_{N}I_{q}$,
where $R_{N}=\rho_{N}\lambda_{N}/A$ and A is the cross sectional area
of N. $\delta\mu$ decays away from the injection point at $x=0$
as $\delta\mu=\delta\mu_{0}\exp(-|x|/\lambda_{N})$, exponentially on
the scale of $\lambda_{N}$. F detector electrodes can be placed at
$x=L_{1},\  L_{2},\ ...$, near the injection point. The voltage induced
over the detector junction is $V(L)=\delta\mu(L)\gamma/2e$, and the
{}``spin signal'' \cite{Jedema:02,Urech:04,Valenzuela:05} becomes
\begin{equation}
R_{S}=\frac{V}{I_{inj}}=\gamma^{2}R_{N}\exp{(-L/\lambda_{N})}.\label{eq:1}\end{equation}

At small injection energies, of the order of the superconducting gap
energy, and currents smaller than the critical current, spin injection
into S via tunnel junctions is expected to populate the quasiparticle
band while the accompanying charge is transferred to the Cooper
pairs. A Cooper pair combines two electrons of opposite spin and therefore
is unable to carry any spin current. Spin current is carried by QP's that
do not necessarily carry charge. Injection of electrons into S can
also result in Andreev reflections \cite{Andreev:64}; a process
that is important for metallic contacts. However, for F-I-S injection the
QP-creation process rather than the Andreev processes is dominant.
The QP's diffuse away from the injection point while their spin relaxes.
This diffusion process is characterized by $D_{S}$ and $\tau_{S}$.
A recent theory \cite{Takahashi:03} has predicted a large enhancement of spin
accumulation in S due to a combined effect of slower QP diffusion,
longer $\tau_{S}$, and a decrease in the QP population by the opening of the
gap in S. 

The samples were made using a two-angle deposition technique, in which
a shadow mask was patterned using a standard electron beam lithography.
10 nm thick Al was e-gun evaporated at normal incidence to form the
Al strip, which was then exposed to oxygen at $80$ mTorr for 15 min
to form a thin oxide layer at the surface. After the oxidation 50
nm thick Cobalt electrodes were deposited at an angle to the Al deposition
to overlap the Al strip. Thus, closely spaced tunnel junctions were
formed in the overlap regions. The junctions were designed to be $\sim150$
nm from the end of the Co electrodes in order to avoid the fringing
magnetic fields from the magnetic tips. We found this design feature
to be important since the fringing fields are estimated to 1-10 kOe at the end 
of the Co electrodes, which can exceed the perpendicular critical field of Al. 
Figure 1 is an SEM micrograph of a typical multi terminal device that we have studied.
It consists of 3 vertical F electrodes numbered 1, 2 and 3 in Fig.
1. The electrodes are $\sim90$, $130$ and $190$ nm wide respectively
and are spaced 280 and 450 nm apart. The Al strip is 150 nm wide and
$40~\mu$m long, with metallic connections at the ends, marked schematically
0 and 4 in Fig. 1. The Co electrodes were made of different width
in order to achieve different coercive fields. This allows for the
total of 8 magnetic states of the device, which can be manipulated
by an external in-plane magnetic field. At 4 K the ferromagnetic junctions
resistances were 20, 14, 10 k$\Omega$. From the conductivity of the
Al strip, $1.3\times10^{7}~\Omega^{-1}$m$^{-1}$, we determine $D_{N}=3.4\times10^{-3}$
m$^{2}$s$^{-1}$ from the Einstein relation $\sigma=e^{2}N_{Al}D_{N}$,
with a density of states at the Fermi level of $N_{Al}=2.4 \times 10^{28}$ 
states per eV per m$^3$ \cite{Jedema:02}. The bias current $I_{14}$ was sent 
between electrodes 1 and 4. The voltages 
were measured between the Al strip at 0 and the Co electrodes 2 and
3, $V_{02}$ and $V_{03}$, using a standard lockin technique. Preamplifiers
with very high input impedance ($\sim10^{13}~\Omega$) and low input
bias currents ($\sim1$ pA) were used in order to minimize possible
artifacts due to the detection circuit. 

Figure 2 shows the non-local spin signal $R_S=V_{20}/I_{14}$ for a field
sweep at $T=0.25$ K and $I_{14}=10~\mu$A. At this high bias current
the spin signal, $R_{S}=0.18~\Omega$, was equal to that at $T=4$
K, where the Al is in the normal state. The switching fields were
$\sim2$ and $\sim1$ kOe for electrode 1 and 2, respectively. According to the theory of spin diffusion in a 1D channel \cite{Johnson:03} the ratio of the two spin voltages is $V_{20}/V_{30} = \exp(d_{23}/\lambda_N)$ where $d_{23}$ is the distance between the two electrodes. See Reference \cite{Urech:04} for more details on the procedure. 
In the normal state of the wire, the spin diffusion length and the spin 
injection efficiency were determined to be $\lambda_{N}=650$ nm and
$\gamma=12\%$, respectively. Using the normal state conductivity
of Al and $\lambda_{N}$, the spin relaxation time is estimated to
be $\tau_{N}\approx100$ ps. The slight curvature in $V_{02}$ at
zero field is due to the tilting magnetization of the electrode in
the junction area.

It is important to verify that changing from the parallel to the antiparallel
state of the injector-detector pair does not affect the superconducting
properties of the Al. Figure 3 shows the superconducting
critical current of the Al strip, $I_{C}$, as a function of the bias
current measured as shown in the inset to Fig. 3. High $I_{bias}$
produces a high population of QP's in the superconductor with a zero
net charge but with either a zero (parallel state) or non-zero (anti-parallel
state) net spin. One can see that as the rate of the QP injection
increases, $I_{C}$ gradually decreases to zero at $I_{bias}=125$ nA, which means
that the Al strip becomes resistive in the injection region. A finite electric field appears due to the so-called phase slips in the superconductor \cite{SBT}. The Al strip is thus resistive, though still appreciably superconducting.  $I_C$ was the same within $\sim 3\%$ for both polarities of $I_{bias}$. Note also that 
$I_{bias}\ll I_C$. This is to be expected for high QP populations \cite{Schmidt}. 
One can clearly see that $I_{C}$ is independent on the orientation of the Co 
electrodes. This means that neither the fringing fields or the spin accumulation 
in Al suppress the superconducting gap. 

In the superconducting state we observe a reduction in the spin diffusion length, 
$\lambda_S= 300$ nm with a dramatic increase in the
spin signal for low current bias at $T=0.25$ K, as shown in Fig. 4. At
large bias currents the spin signal is equal to that of the normal
state of the wire, but increases by $\sim1000$ times at low bias. To our
knowledge this is the first direct measurement of the non-equilibrium spin accumulation 
of quasiparticles in a superconductor. The large enhancement
in $\Delta\mu$ we observe is qualitatively consistent with the prediction
by Takahashi et al \cite{Takahashi:03}.

We have shown that spin injection via tunnel junctions into a superconductor
leads to a much larger spin accumulation at low bias currents while
for large injection currents $R_{S}$ converges to the normal state
value. $R_{S}$ is large as long as the spin accumulation is a small fraction of 
the gap energy, $\Delta\mu=R_{S} e I \ll \Delta$, so the
superconducting gap is unaffected. The enhancement of $R_{S}$ is
interpreted as due to a decrease in the QP population by the opening of the 
superconducting gap. 

M. U., J. J. and N. P. gratefully acknowledge support from the Swedish SSF
under the Framework Grant for Magnetoelectronic Nanodevice Physics.

\begin{figure}
\centering
\includegraphics[width=1.0\textwidth]{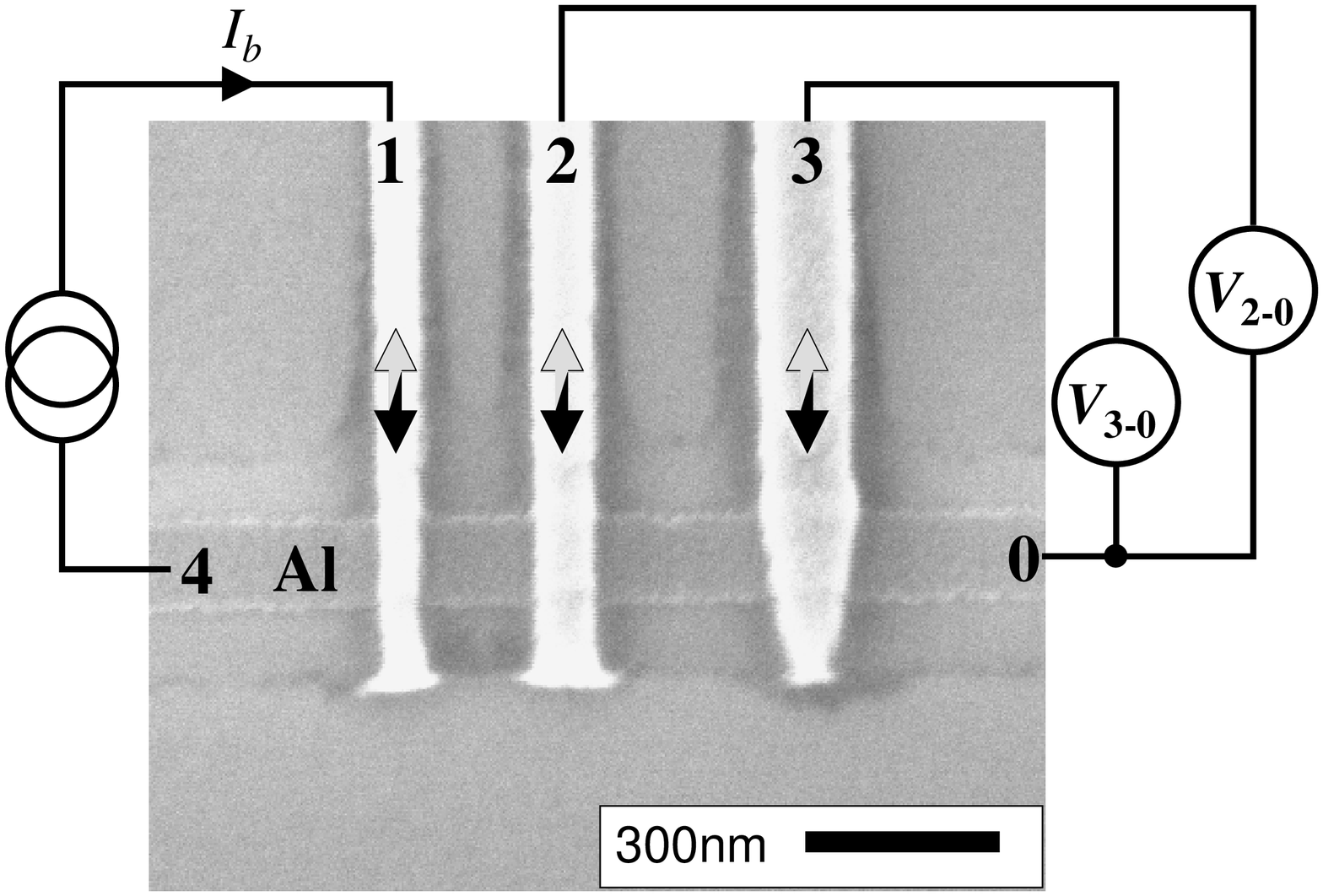}
\caption{Scanning electron micrograph of a multi terminal device. }
\end{figure}

\begin{figure}
\centering
\includegraphics[width=1.0\textwidth]{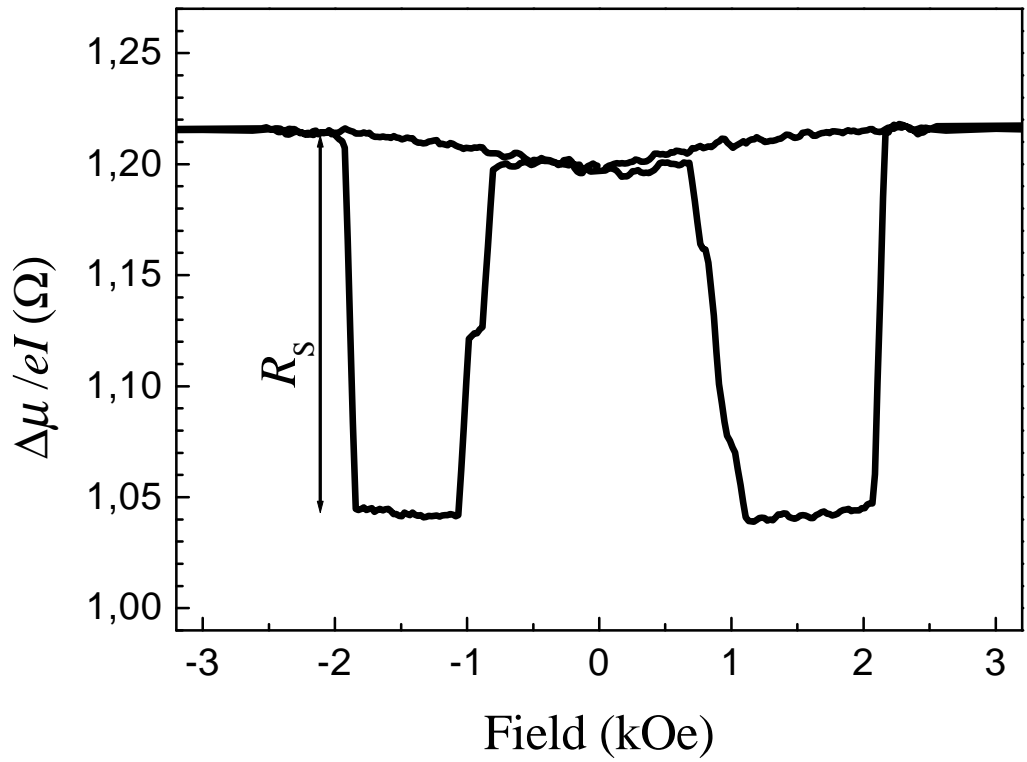}
\caption{Normalized detector signal, $V_{20}/I_{14}$, versus field at 0.25 K. The bias current is $10~\mu$A. The difference between the antiparallel and the parallel
state is proportional to the spin accumulation.}
\end{figure}

\begin{figure}
\centering
\includegraphics[width=1.0\textwidth]{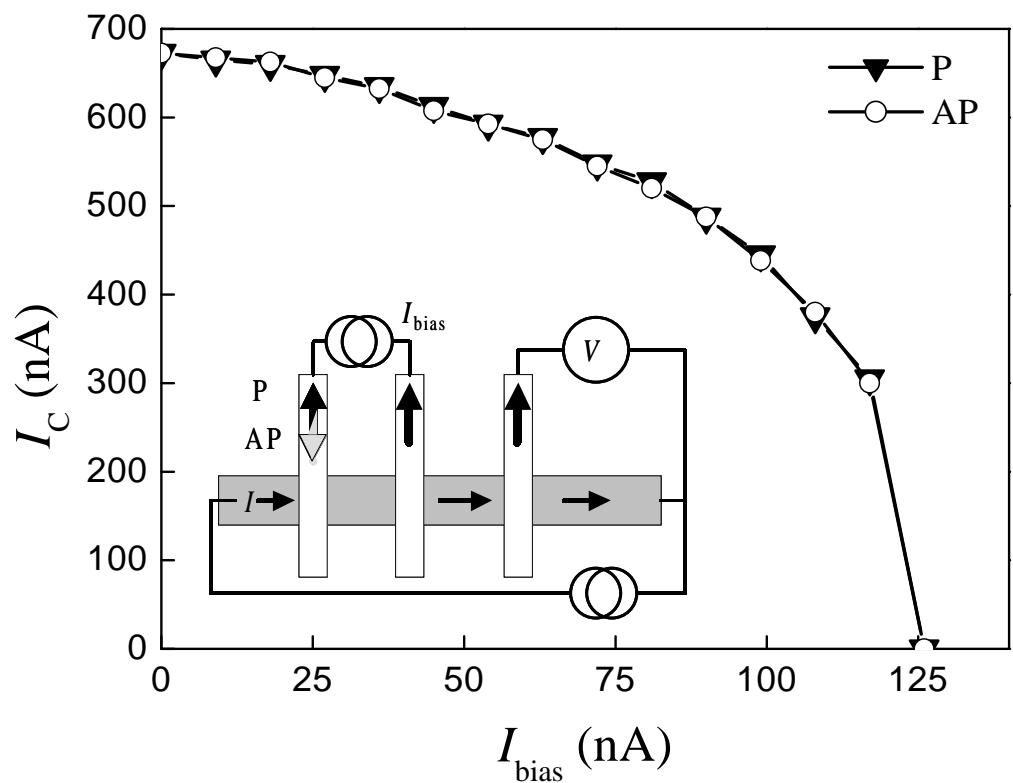}
\caption{Critical current for the Al strip versus $I_{bias}$. The inset shows
the measurement configuration. }
\end{figure}

\begin{figure}
\centering
\includegraphics[width=1.0\textwidth]{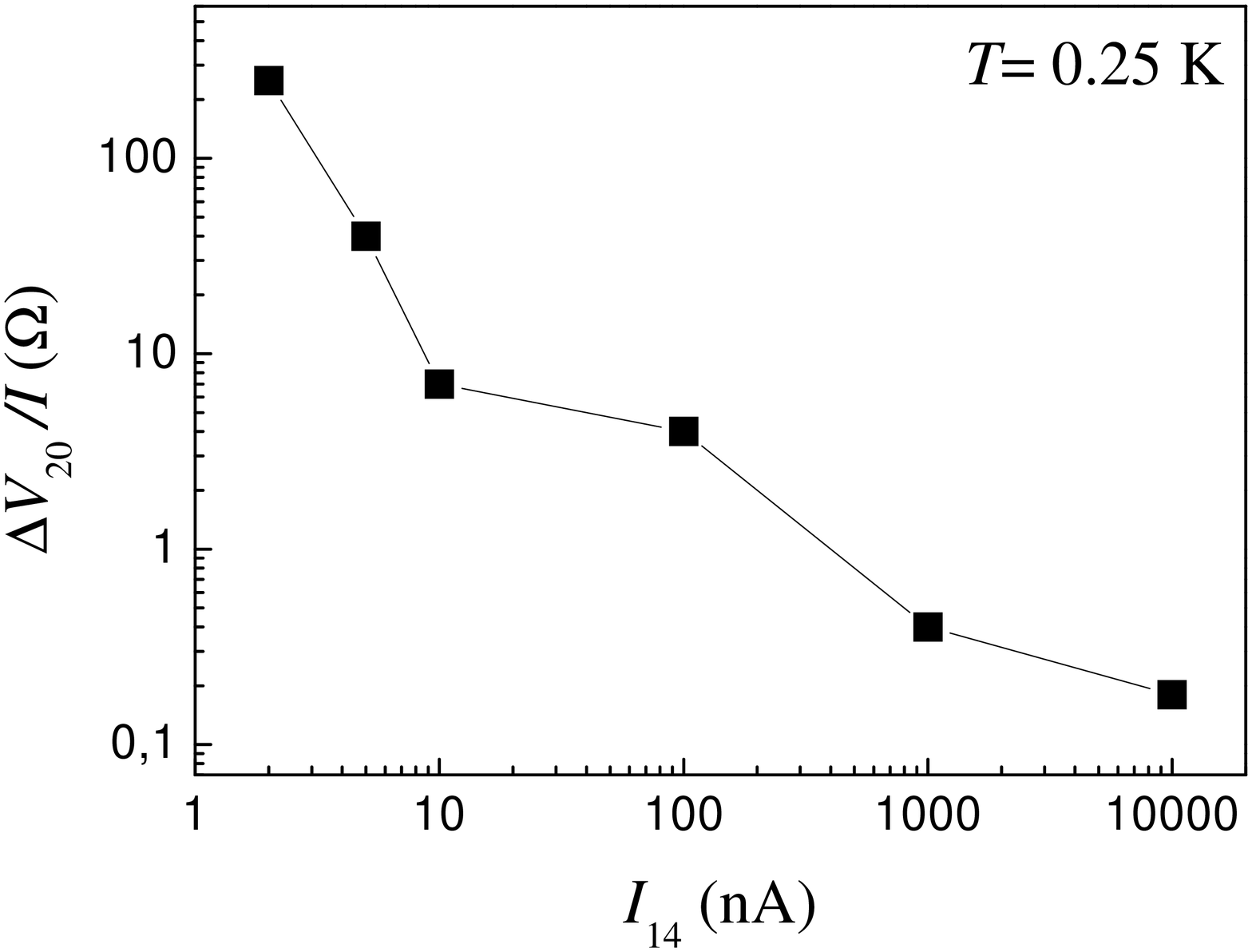}
\caption{Spin signal as a function of bias.}
\end{figure}


\newpage

\begin{thebibliography}{14}
\bibitem{Clarke:72} J. Clarke, Phys. Rev. Lett., \textbf{28}, 1363 (1972) 
\bibitem{Johnson:94}M. Johnson, Appl. Phys. Lett., \textbf{65}, 1460 (1994) 
\bibitem{Gu:02}J.Y. Gu, J. A. Caballero, R. D. Slater, R. Loloee, and W. P. Pratt, Jr , Phys. Rev. B, \textbf{66}, 140507R (2002) 
\bibitem{Shin:03}Y.-S. Shin, Phys. Rev. B, \textbf{71}, 144513 (2005) 
\bibitem{Andreev:64}A. F. Andreev, Sov. Phys. JETP, \textbf{19}, 1228 (1964)
\bibitem{Tedrow:73}P. M. Tedrow and R. Meservey, Phys. Rev. B \textbf{7}, 318 (1973).
\bibitem{Johnson:85}M. Johnson and R. H. Silsbee, Phys. Rev. Lett., \textbf{55}, 1790 (1985)
\bibitem{Jedema:02}F. J. Jedema H. B. Heersche, A. T. Filip, J. J. S. Baselmans, and B. J. van Wees, Nature, \textbf{416},713 (2002) 
\bibitem{Urech:04}M. Urech, J. Johansson, V. Korenivski, and D. B. Haviland, J. Magn. Magn. Mater., \textbf{272-276}, e1417 (2004)
\bibitem{Valenzuela:05}S. O. Valenzuela, D. J. Monsma, C. M. Marcus, V. Narayanamurti and M. Tinkham, Phys. Rev. Lett., \textbf{94}, 196601 (2005) 
\bibitem{Takahashi:03}S. Takahashi and S. Maekawa, Phys. Rev. B, \textbf{67}, 052409 (2003)
\bibitem{Johnson:03}M. Johnson and J. Byers, Phys. Rev. B, \textbf{67}, 125112 (2003) 
\bibitem{SBT}W. Skocpol, M. R. Beasley, and M. Tinkham, J. Low. Temp. Phys., \textbf{16}, 145 (1974)
\bibitem{Schmidt}V. V. Schmidt, The Physics of Superconductors. Eds. P. Muller, A. V. Ustinov. Springer-Verlag, Berlin-Heidelberg (1997)
 
\end{thebibliography}
\end{document}